\documentclass[10pt,conference]{IEEEtran}
\usepackage{cite,graphicx,psfrag,amsmath,amssymb,url}

\def\mid{{\mathop{\rm mid}}}
\def\midt{{\mathop{\rm mid}}}
\def\hi{{\mathop{\rm hi}}}
\def\hit{{\mathop{\rm hi}}}
\def\lo{{\mathop{\rm lo}}}

\def\os{{\mathop{\rm CC}}}
\def\argmin{{\mathop{\arg\,\min}}}
\def\order{O}

\def\for{\mbox{ for }}
\def\st{\mbox{ such that }}
\def\total{{\mathop{\rm tot}}}
\def\power{{\mathop{\rm pow}}}
\def\rhot{{\rho_{\total}}}
\def\rhop{{\rho_{\power}}}

\def\AlphaSet{{A}}
\def\BetaSet{{B}}
\def\GammaSet{{\mathit{\Gamma}}}
\def\DeltaSet{{\mathit{\Delta}}}
\def\algd{D}
\def\nodeset{\eta}

\def\definedas{\triangleq}
\def\das{\triangleq}
\def\smalll{{\mbox{\scriptsize \boldmath $l$}}}
\def\boldl{{\mbox{\boldmath $l$}}}
\def\letterl{l}

\def\p{{\mbox{\boldmath $p$}}}

\def\I{{\mathcal I}}

\def\letterk{{\kappa}}
\def\P{{\mathcal P}}
\def\PS{{\mathcal P}^*}
\def\R{{\mathbb R}}
\def\Rp{{\mathbb R}_+}
\def\S{{\mathcal S}}

\def\X{{\mathcal X}}
\def\Z{{\mathbb Z}}

\def\boldlupN{{\boldl^{N}}}
\newcommand{\defn}[0]{\textit}
\newtheorem{theorem}{Theorem}

\newtheorem{definition}{Definition}

\begin{document}
\title{$D$-ary Bounded-Length Huffman Coding}

\author{\authorblockN{Michael B. Baer}
\authorblockA{Electronics for Imaging\\
303 Velocity Way\\
Foster City, California  94404  USA\\
Email: Michael.Baer@efi.com}}

\maketitle

\begin{abstract}
Efficient optimal prefix coding has long been accomplished via the Huffman algorithm.  However, there is still room for improvement and exploration regarding variants of the Huffman problem.  Length-limited Huffman coding, useful for many practical applications, is one such variant, in which codes are restricted to the set of codes in which none of the $n$ codewords is longer than a given length, $l_{\max}$. Binary length-limited coding can be done in $O(n l_{\max})$ time and $O(n)$ space via the widely used Package-Merge algorithm.  In this paper the Package-Merge approach is generalized without increasing complexity in order to introduce a minimum codeword length, $l_{\min}$, to allow for objective functions other than the minimization of expected codeword length, and to be applicable to both binary and nonbinary codes; nonbinary codes were previously addressed using a slower dynamic programming approach.  These extensions have various applications --- including faster decompression --- and can be used to solve the problem of finding an optimal code with limited fringe, that is, finding the best code among codes with a maximum difference between the longest and shortest codewords.  The previously proposed method for solving this problem was nonpolynomial time, whereas solving this using the novel algorithm requires only $O(n (l_{\max}- l_{\min})^2)$ time and $O(n)$ space.
\end{abstract}

\section{Introduction} 
\label{intro} 
A source emits input symbols drawn from the alphabet $\X = \{ 1, 2,
\ldots, n \}$, where $n$ is an integer.  Symbol $i$ has probability
$p_i$, thus defining probability vector $\p=(p_1, p_2, \ldots, p_n)$.
Only possible symbols are considered for coding and these are, without
loss of generality, sorted in decreasing order of probability; thus
$p_i > 0$ and $p_i \leq p_j$ for every $i>j$ such that $i, j \in \X$.
Each input symbol is encoded into a codeword composed of output
symbols of the $\algd$-ary alphabet $\{0, 1, \ldots, \algd - 1\}$.
The codeword $c_i$ corresponding to input symbol $i$ has length
$\letterl_i$, thus defining length vector $\boldl = (l_1, l_2, \ldots,
l_n)$.  The code should be a prefix code, i.e., no codeword
$c_i$ should begin with the entirety of another codeword~$c_j$. 

For the \defn{bounded-length} coding variant of Huffman coding
introduced here, all codewords must have lengths lying in a given
interval [$l_{\min}$,$l_{\max}$].  Consider an application in the
problem of designing a data codec which is efficient in terms of both
compression ratio and coding speed.  Moffat and Turpin proposed a
variety of efficient implementations of prefix encoding and decoding
in \cite{MoTu97}, each involving table lookups rather than code trees.
They noted that the length of the longest codeword should be limited
for computational efficiency's sake.  Computational efficiency is also
improved by restricting the overall range of codeword lengths,
reducing the size of the tables and the expected time of searches
required for decoding.  Thus, one might wish to have a minimum
codeword size of, say, $l_{\min}=16$ bytes and a maximum codeword size
of $l_{\max}=32$ bytes ($\algd=2$).  If expected codeword length 
for an optimal code found under these restrictions is too long,
$l_{\min}$ can be reduced and the algorithm rerun until the proper
trade-off between coding speed and compression ratio is found.

A similar problem is one of determining opcodes of a
microprocessor designed to use variable-length opcodes, each a certain
number of bytes ($\algd=256$) with a lower limit and an upper limit to
size, e.g., a restriction to opcodes being 16, 24, or 32 bits long
($l_{\min}=2$, $l_{\max}=4$).  This problem clearly falls within the
context considered here, as does the problem of assigning video
recorder scheduling codes; these human-readable decimal codes
($\algd=10$) also have bounds on their size, such as
$l_{\min} = 3$ and $l_{\max} = 8$.

Other problems of interest have $l_{\min} = 0$ and are thus length
limited but have no practical lower bound on length\cite[p.~396]{WMB}.
Yet other problems have not fixed bounds but a constraint on the
difference between minimum and maximum codeword length, a quantity
referred to as fringe \cite[p.~121]{Abr01}.  As previously noted,
large fringe has a negative effect of the speed of a decoder.  

If we either do not require a minimum or do not require a maximum, it
is easy to find values for $l_{\min}$ or $l_{\max}$ that do not limit
the problem.  As mentioned, setting $l_{\min} = 0$ results in a
trivial minimum, as does $l_{\min} = 1$.  Similarly, setting $l_{\max}
= n$ or using the hard upper bound $l_{\max} = \lceil (n-1)/(\algd-1)
\rceil$ results in a trivial maximum value.  

If both minimum and maximum values are trivial, Huffman
coding~\cite{Huff} yields a prefix code minimizing expected codeword
length $\sum_i p_i \letterl_i$.  The conditions necessary and
sufficient for the existence of a prefix code with length vector
$\boldl$ are the integer constraint, $\letterl_i \in \Z_+$, and the
Kraft (McMillan) inequality~\cite{McMi},
\begin{equation}
\letterk(\boldl) \definedas \sum_{i=1}^n \algd^{-\letterl_i}\leq 1 .
\label{kraft}
\end{equation}
Finding values for $\boldl$ is
sufficient to find a corresponding code.

It is not always obvious that we should minimize the expected number
of questions $\sum_i p_i l_i$ (or, equivalently, the expected number
of questions in excess of the first~$l_{\min}$, $\sum_i p_i
(l_i-l_{\min})^+$, where $x^+$ is $x$ if $x$ is positive, $0$
otherwise).  We generalize and investigate how to minimize the value
\begin{equation}
\sum_{i=1}^n p_i \varphi(l_i-l_{\min})
\label{penalty}
\end{equation}
under the above constraints
for any \defn{penalty function} $\varphi(\cdot)$ convex and increasing
on~$\Rp$.  Such an additive measurement of cost is called a
\defn{quasiarithmetic penalty}, in this case a convex quasiarithmetic
penalty.

One such function $\varphi$ is $\varphi(\delta) = (\delta+l_{\min})^2$, a
quadratic value useful in optimizing a communications delay
problem~\cite{Larm}.  Another function, $\varphi(\delta) =
\algd^{t(\delta+l_{\min})}$ for $t>0$, can be used to minimize the
probability of buffer overflow in a queueing system\cite{Humb2}.

Mathematically stating the bounded-length problem,
$$
\begin{array}{ll}
\mbox{Given } & \p = (p_1, \ldots, p_n),~p_i > 0; \\
& \algd \in \{2, 3, \ldots\}; \\
& \mbox{convex, monotonically increasing } \\
& \varphi: \Rp \rightarrow \Rp \\
\mbox{Minimize} \,_{\{\smalll\}} &
\sum_i p_i \varphi(l_i-l_{\min}) \\
\mbox{subject to } & \sum_i \algd^{-l_i} \leq 1; \\
& l_i \in \{l_{\min},l_{\min}+1,\ldots,l_{\max}\}.
\end{array}
$$ Note that we need not assume that probabilities $p_i$ sum to $1$;
they could instead be arbitrary positive weights.  

Given a finite $n$-symbol input alphabet with an associated
probability vector $\p$, a $\algd$-symbol output alphabet with
codewords of lengths $[l_{\min},l_{\max}]$ allowed, and a
constant-time calculable penalty function $\varphi$, we describe an
$\order(n(l_{\max}-l_{\min}))$-time $\order(n)$-space algorithm for
constructing a $\varphi$-optimal code.  In Section~\ref{prelim},
we present a brief review of the relevant literature before extending
to $\algd$-ary codes a notation first presented in \cite{Larm}.  This
notation aids in solving the problem in question by reformulating it
as an instance of the $\algd$-ary Coin Collector's problem, presented
in the section as an extension of the original (binary) Coin
Collector's problem\cite{LaHi}.  An extension of the Package-Merge
algorithm solves this problem; we introduce the reduction and
resulting algorithm in Section~\ref{algorithm}.  An application to a
previously proposed problem involving tree fringe is discussed in
Section~\ref{conclusion}.

\section{Preliminaries}
\label{prelim}

Reviewing how the problem in question differs from binary Huffman
coding:
\begin{enumerate}
\item It can be nonbinary, a case considered by Huffman in his original
paper\cite{Huff};
\item There is a maximum codeword length, a restriction previously
considered, e.g., \cite{Itai} in $\order(n^3 l_{\max} \log \algd)$
time \cite{GoWo} and $\order(n^2 \log \algd)$ space, but solved
efficiently only for binary coding, e.g., \cite{LaHi} in $\order(n
l_{\max})$ time $\order(n)$ space and most efficiently in \cite{Schi};
\item There is a minimum codeword length, a novel restriction;
\item The penalty can be nonlinear, a modification previously
considered, but only for binary coding, e.g., \cite{Baer06}.
\end{enumerate}

The minimum size constraint on codeword length requires a relatively
simple change of solution range to \cite{LaHi}.  The nonbinary coding
generalization is a bit more involved; it requires first modifying the
Package-Merge algorithm to allow for an arbitrary numerical base
(binary, ternary, etc.), then modifying the coding problem to allow for
a provable reduction to the modified Package-Merge algorithm.  
The $\order(n(l_{\max}-l_{\min}))$-time $\order(n)$-space algorithm
minimizes \defn{height} (that is, maximum codeword length) among
optimal codes (if multiple optimal codes exist).

Before presenting an algorithm for optimizing the above problem, we
introduce a notation for codes that generalizes one first presented
in~\cite{Larm} and modified in \cite{Baer06}.  

\textit{The key idea:} Each node $(i,l)$ represents both the share of
the penalty (\ref{penalty}) (\defn{weight}) and the (scaled) share of
the Kraft sum (\ref{kraft}) (\defn{width}) assumed for the $l$th bit
of the $i$th codeword.  By showing that total weight is an increasing
function of the penalty and that there is a one-to-one correspondence
between an optimal code and a corresponding optimal nodeset, we reduce
the problem to an efficiently solvable problem, the Coin Collector's
problem.

In order to do this, we first need to make a modification to the
problem analogous to one Huffman made in his original nonbinary
solution.  We must in some cases add a ``dummy'' input or ``dummy''
inputs of probability $p_i = 0$ to the probability vector to assure
that the optimal code has the Kraft inequality satisfied with
equality, an assumption underlying both the Huffman algorithm and
ours.  If we use the minimum number of dummy inputs needed to make $n
\bmod{(\algd-1)} \equiv 1$, we can assume without loss of generality
that $\letterk(\boldl)=1$.  With this modification, we present
\defn{nodeset} notation:

\begin{definition} A \defn{node} is an ordered pair of integers $(i, l)$ such
  that $i \in \{1,\ldots,n\}$ and $l \in
  \{l_{\min}+1,\ldots,l_{\max}\}$.  Call the set of all possible
  nodes~$I$.  This set can be arranged in an $n \times
  (l_{\max}-l_{\min})$ grid, e.g., Fig.~\ref{nodesetnum}.  The set of
  nodes, or \defn{nodeset}, corresponding to input symbol $i$ (assigned
  codeword~$c_i$ with length~$l_i$) is the set of the
  first~$l_i-l_{\min}$ nodes of column~$i$, that is,
  $\nodeset_\smalll(i) \definedas \{(j,l)~|~j=i,~l \in
  \{l_{\min}+1,\ldots,l_i\}\}$.  The nodeset corresponding to length
  vector~$\boldl$ is $\nodeset(\boldl) \definedas \bigcup_i
  \nodeset_\smalll(i)$; this corresponds to a set of $n$ codewords, a
  code.  Thus, in Fig.~\ref{nodesetnum}, the dashed line surrounds a
  nodeset corresponding to $\boldl = (1, 2, 2, 2, 2, 2, 2)$.  We say a
  node $(i,l)$ has \defn{width} $\rho(i,l) \definedas \algd^{-l}$ and
  \defn{weight} $\mu(i,l) \definedas p_i \varphi(l-l_{\min}) - p_i
  \varphi(l-l_{\min}-1)$, as shown in the example in Fig.~\ref{nodesetnum}.
  Note that if $\varphi(l)=l$, $\mu(i,l)=p_i$.
\end{definition}
Given valid nodeset $N \subseteq I$, it is straightforward to find the
corresponding length vector and, if it satisfies the Kraft inequality,
a code.  

We find an optimal nodeset using the $\algd$-ary Coin
Collector's problem.  Let $\algd^\Z$ denote the set of all integer
powers of a fixed integer $\algd>1$.  The Coin Collector's problem of
size $m$ considers ``coins'' indexed by $i \in \{1, 2, \ldots, m\}$.
Each coin has a width, $\rho_i \in \algd^\Z$; one can think of width
as coin face value, e.g., $\rho_i = 0.25 = 2^{-2}$ for a quarter
dollar (25 cents).  Each coin also has a weight, $\mu_i \in \R$.  The
final problem parameter is total width, denoted~$\rhot$.  The problem
is then:
\begin{equation}
\begin{array}{ll}
\mbox{Minimize} \,_{\{B \subseteq \{1,\ldots,m\}\}} & \sum_{i \in B} \mu_i  \\
\mbox{subject to } & \sum_{i \in B} \rho_i = \rhot \\
\mbox{where } & m \in \Z_+ , \mu_i \in \R \\
& \rho_i \in \algd^\Z , \rhot \in \R_+ .
\end{array} \label{knap}
\end{equation}
We thus wish to choose coins with total width $\rhot$ such that their
total weight is as small as possible.  This problem has a linear-time solution
given sorted inputs; this solution was found for $\algd=2$ in \cite{LaHi} and is found for $\algd>2$ here. 

Let $i \in \{1,\ldots,m\}$ denote both the index of a coin and the
coin itself, and let $\I$ represent the $m$ items along with their
weights and widths.  The optimal solution, a function of total width
$\rhot$ and items $\I$, is denoted $\os(\I,\rhot)$ (the optimal coin
collection for $\I$ and $\rhot$).  Note that, due to ties, this need
not be a unique solution, but the algorithm proposed here is
deterministic; that is, it finds one specific solution, much like
bottom-merge Huffman coding\cite{Schw} or the corresponding
length-limited problem\cite{LaPr2,Baer06}

Because we only consider cases in which a solution exists, $\rhot =
\omega \rhop$ for some $\rhop \in \algd^\Z$ and $\omega \in \Z_+$.
Here, assuming $\rhot>0$, $\rhop$ and $\omega$ are the unique pair of
a power of $\algd$ and an integer that is not a multiple of $\algd$,
respectively, which, multiplied, form~$\rhot$.  If $\rhot = 0$,
$\omega$ and $\rhop$ are not used.  Note that $\rhop$ need not be an
integer.

${}$

\noindent \textbf{Algorithm variables} \\
At any point in the algorithm, given nontrivial $\I$ and $\rhot$, we use the following definitions: 

\begin{tabular}{rcl}
Remainder & & \\
$\rhop$ & $\definedas$ & the unique $x \in \algd^\Z$ \\
& & such that $\frac{\rhot}{x} \in \Z \backslash \algd\Z$ \\[2pt]
Minimum width & & \\
$\rho^*$ & $\definedas$ & $\min_{i \in \I} \rho_i$ ($\in \algd^\Z$)\\[2pt]
Small width set & & \\
$\I^*$ & $\definedas$ & $\{i~|~\rho_i = \rho^*\}$ ($\neq \emptyset$)\\[2pt]
``First'' item & & \\
$i^*$ & $\definedas$ & $\argmin_{i \in \I^*} \mu_i$ \\
& & (ties broken w/highest index) \\[2pt]
``First'' package & & \\
$\PS$ & $\definedas$ & 
$\left\{\begin{array}{ll}
\P \st & \\
\quad |\P|=\algd, & \\
\quad \P \subseteq \I^*, & \\
\quad \P \preceq \I^* \backslash \P, & |\I^*| \geq \algd \\
\emptyset, & |\I^*| < \algd \\
\end{array}
\right.$ \\
& & (ties broken w/highest indices) \\[2pt]
\end{tabular}

${}$

\noindent where $\algd \Z$ denotes integer multiples of $\algd$ and $\P \preceq
\I^* \backslash \P$ denotes that, for all $i \in \P$ and $j \in \I^*
\backslash \P$, $\mu_i \leq \mu_j$.  Then the following is a recursive
description of the algorithm:

${}$

\noindent \textbf{Recursive $\algd$-ary Package-Merge Procedure}

\textit{Basis.  $\rhot = 0$}:  $\os (\I,\rhot)=\emptyset$.

\textit{Case 1.  $\rho^* = \rhop$ and $\I \neq \emptyset$}:  $\os(\I,\rhot) =
\os (\I \backslash \{i^*\},\rhot-\rho^*) \cup \{i^*\}$.

\textit{Case 2a.  $\rho^* < \rhop$, $\I \neq \emptyset$, and $|\I^*| < \algd$}:
$\os(\I,\rhot) = \os(\I \backslash \I^*, \rhot)$.

\textit{Case 2b.  $\rho^* < \rhop$, $\I \neq \emptyset$, and $|\I^*|
\geq \algd$}: Create $i'$, a new item with weight $\mu_{i'} =
\sum_{i\in \PS} \mu_i$ and width $\rho_{i'} = \algd \rho^*$.  This new
item is thus a combined item, or \defn{package}, formed by combining
the $\algd$ least weighted items of width~$\rho^*$.  Let $\S = \os(\I
\backslash \PS \cup \{i'\},\rhot)$ (the optimization of the packaged
version, where the package is given a low index so that, if
``repackaged,'' this occurs after all singular or previously packaged
items of identical weight and width).  If $i' \in \S$, then
$\os(\I,\rhot) = \S \backslash \{i'\} \cup \PS$; otherwise,
$\os(\I,\rhot) = \S$.

\begin{theorem}
If an optimal solution to the Coin Collector's problem exists, the
above recursive (Package-Merge) algorithm will terminate with an
optimal solution. 
\end{theorem}

\begin{proof}
Using induction on the number of input items, while the basis is
trivially correct, each inductive case reduces the number of items by
at least one.  The inductive hypothesis on $\rhot \geq 0$ and $\I \neq
\emptyset$ is that the algorithm is correct for any problem instance
with fewer input items than instance $(\I,\rhot)$.

If $\rho^* > \rhop > 0$, or if $\I=\emptyset$ and $\rhot \neq 0$, then
there is no solution to the problem, contrary to our assumption.  Thus
all feasible cases are covered by those given in the procedure.  Case
1 indicates that the solution must contain at least one element (item
or package) of width~$\rho^*$.  These must include the minimum weight
item in $\I^*$, since otherwise we could substitute one of the items
with this ``first'' item and achieve improvement.  Case~2 indicates
that the solution must contain a number of elements of width $\rho^*$
that is a multiple of~$\algd$.  If this number is $0$, none of the
items in $\PS$ are in the solution.  If it is not, then they all are.
Thus, if $\PS = \emptyset$, the number is $0$, and we have Case 2a.
If not, we may ``package'' the items, considering the replaced package
as one item, as in Case 2b.  Thus the inductive hypothesis holds.
\end{proof}

The algorithm can be performed in linear time and space, as with the
binary version \cite{LaHi}.

\section{A General Algorithm}
\label{algorithm}

\begin{figure*}[ht]
\centering
\psfrag{l (level)}{$l$ (level)}
\psfrag{i (item)}{$i$ (input symbol)}
\psfrag{(width)}{$\rho$ (width)}
\psfrag{1m1}{\scriptsize $\mu(1,2) = p_1$}
\psfrag{2m1}{\scriptsize $\mu(2,2) = p_2$}
\psfrag{3m1}{\scriptsize $\mu(3,2) = p_3$}
\psfrag{4m1}{\scriptsize $\mu(4,2) = p_4$}
\psfrag{5m1}{\scriptsize $\mu(5,2) = p_5$}
\psfrag{6m1}{\scriptsize $\mu(6,2) = p_6$}
\psfrag{7m1}{\scriptsize $\mu(7,2) = p_7$}
\psfrag{1m2}{\scriptsize $\mu(1,3) =3p_1$}
\psfrag{2m2}{\scriptsize $\mu(2,3) =3p_2$}
\psfrag{3m2}{\scriptsize $\mu(3,3) =3p_3$}
\psfrag{4m2}{\scriptsize $\mu(4,3) =3p_4$}
\psfrag{5m2}{\scriptsize $\mu(5,3) =3p_5$}
\psfrag{6m2}{\scriptsize $\mu(6,3) =3p_6$}
\psfrag{7m2}{\scriptsize $\mu(7,3) =3p_7$}
\psfrag{1m3}{\scriptsize $\mu(1,4) =5p_1$}
\psfrag{2m3}{\scriptsize $\mu(2,4) =5p_2$}
\psfrag{3m3}{\scriptsize $\mu(3,4) =5p_3$}
\psfrag{4m3}{\scriptsize $\mu(4,4) =5p_4$}
\psfrag{5m3}{\scriptsize $\mu(5,4) =5p_5$}
\psfrag{6m3}{\scriptsize $\mu(6,4) =5p_6$}
\psfrag{7m3}{\scriptsize $\mu(7,4) =5p_7$}
\psfrag{1r1}{\scriptsize $\rho(1,2) = \frac{1}{9}$}
\psfrag{1r2}{\scriptsize $\rho(2,2) = \frac{1}{9}$}
\psfrag{1r3}{\scriptsize $\rho(3,2) = \frac{1}{9}$}
\psfrag{1r4}{\scriptsize $\rho(4,2) = \frac{1}{9}$}
\psfrag{1r5}{\scriptsize $\rho(5,2) = \frac{1}{9}$}
\psfrag{1r6}{\scriptsize $\rho(6,2) = \frac{1}{9}$}
\psfrag{1r7}{\scriptsize $\rho(7,2) = \frac{1}{9}$}
\psfrag{2r1}{\scriptsize $\rho(1,3) = \frac{1}{27}$}
\psfrag{2r2}{\scriptsize $\rho(2,3) = \frac{1}{27}$}
\psfrag{2r3}{\scriptsize $\rho(3,3) = \frac{1}{27}$}
\psfrag{2r4}{\scriptsize $\rho(4,3) = \frac{1}{27}$}
\psfrag{2r5}{\scriptsize $\rho(5,3) = \frac{1}{27}$}
\psfrag{2r6}{\scriptsize $\rho(6,3) = \frac{1}{27}$}
\psfrag{2r7}{\scriptsize $\rho(7,3) = \frac{1}{27}$}
\psfrag{3r1}{\scriptsize $\rho(1,4) = \frac{1}{81}$}
\psfrag{3r2}{\scriptsize $\rho(2,4) = \frac{1}{81}$}
\psfrag{3r3}{\scriptsize $\rho(3,4) = \frac{1}{81}$}
\psfrag{3r4}{\scriptsize $\rho(4,4) = \frac{1}{81}$}
\psfrag{3r5}{\scriptsize $\rho(5,4) = \frac{1}{81}$}
\psfrag{3r6}{\scriptsize $\rho(6,4) = \frac{1}{81}$}
\psfrag{3r7}{\scriptsize $\rho(7,4) = \frac{1}{81}$}
\psfrag{1}{$1$}
\psfrag{2}{$2$}
\psfrag{3}{$3$}
\psfrag{4}{$4$}
\psfrag{5}{$5$}
\psfrag{6}{$6$}
\psfrag{7}{$7$}
\resizebox{14cm}{!}{\includegraphics{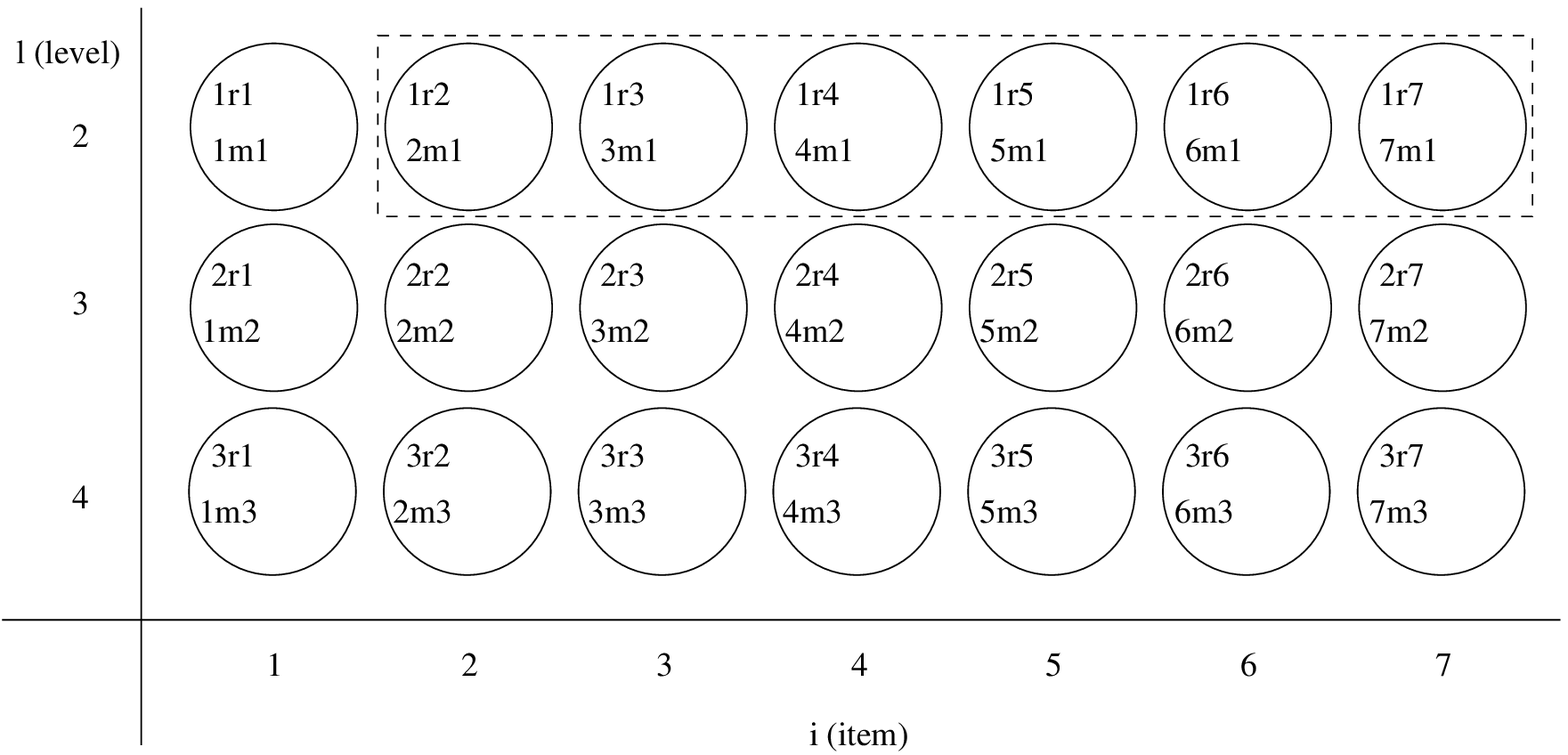}}
\caption{The set of nodes $I$ with widths $\{\rho(i,l)\}$ and weights
$\{\mu(i,l)\}$ for $\varphi(\delta) = \delta^2$, $n=7$, $\algd=3$,
$l_{\min} = 1$, $l_{\max}=4$}
\label{nodesetnum}
\end{figure*}

\begin{theorem}
The solution $N$ of the Package-Merge algorithm for $\I=I$ and $$\rhot
=\frac{n-\algd^{l_{\min}}}{\algd-1}\algd^{-l_{\min}}$$ has a
corresponding length vector $\boldlupN$ such that $N =
\nodeset(\boldlupN)$ and $\mu(N) = \min_{\smalll} \sum_i p_i
\varphi(l_i-l_{\min}) - \varphi(0) \sum_i p_i$.
\end{theorem}

A formal proof can be found in the full version at \cite{Baer20}.  The
idea is to show that, if there is an $(i,l) \in N$ with $l \in
[l_{\min}+2, l_{\max}]$ such that $(i,l-1) \in I \backslash N$, one
can strictly decrease the penalty by substituting item $(i,l-1)$ for a
set of items including $(i,l)$, showing the suboptimality of $N$.
Conversely, if there is no such $(i,l)$, optimal $N$ corresponds to an
optimal length vector.

Because the Coin Collector's problem is linear in time and space ---
same-width inputs are presorted by weight, numerical operations and
comparisons are constant time --- the overall algorithm finds an
optimal code in $\order(|\I|) = \order(n(l_{\max}-l_{\min}))$ time and
space.  Space complexity, however, can be lessened.  This is because
the algorithm output is a monotonic nodeset:

\begin{definition} A \defn{monotonic} nodeset, $N$, is one with the following properties:
\begin{eqnarray}
&(i,l) \in N \Rightarrow (i+1,l) \in N& \for i<n \label{firstprop} \quad \\
&(i,l) \in N \Rightarrow (i,l-1) \in N& \for l>l_{\min}+1 . \quad \label{validlen} 
\end{eqnarray}
In other words, a nodeset is monotonic if and only if it corresponds
to a length vector $\boldl$ with lengths sorted in increasing
order; this definition is equivalent to that given in \cite{LaHi}.
\end{definition}

\begin{figure*}[ht]
\centering
\psfrag{A}{$\AlphaSet$}
\psfrag{B}{$\BetaSet$}
\psfrag{C}{$\GammaSet$}
\psfrag{D}{$\DeltaSet$}
\psfrag{N}{$N$}
\psfrag{3negmin}{$\algd^{-l_{\min}+1}$}
\psfrag{3negmid}{$\algd^{-l_{\midt}}$}
\psfrag{3negmax}{$\algd^{-l_{\max}}$}
\psfrag{lmin}{$l_{\min}+1$}
\psfrag{lmid}{$l_\mid$}
\psfrag{lmax}{$l_{\max}$}
\psfrag{n}{$n$}
\psfrag{n-m}{$n-n_\nu$}
\psfrag{1}{$1$}
\psfrag{l (level)}{$l$ (level)}
\psfrag{i (item)}{$i$ (input symbol)}
\psfrag{(width)}{$\rho$ (width)}
\resizebox{15cm}{!}{\includegraphics{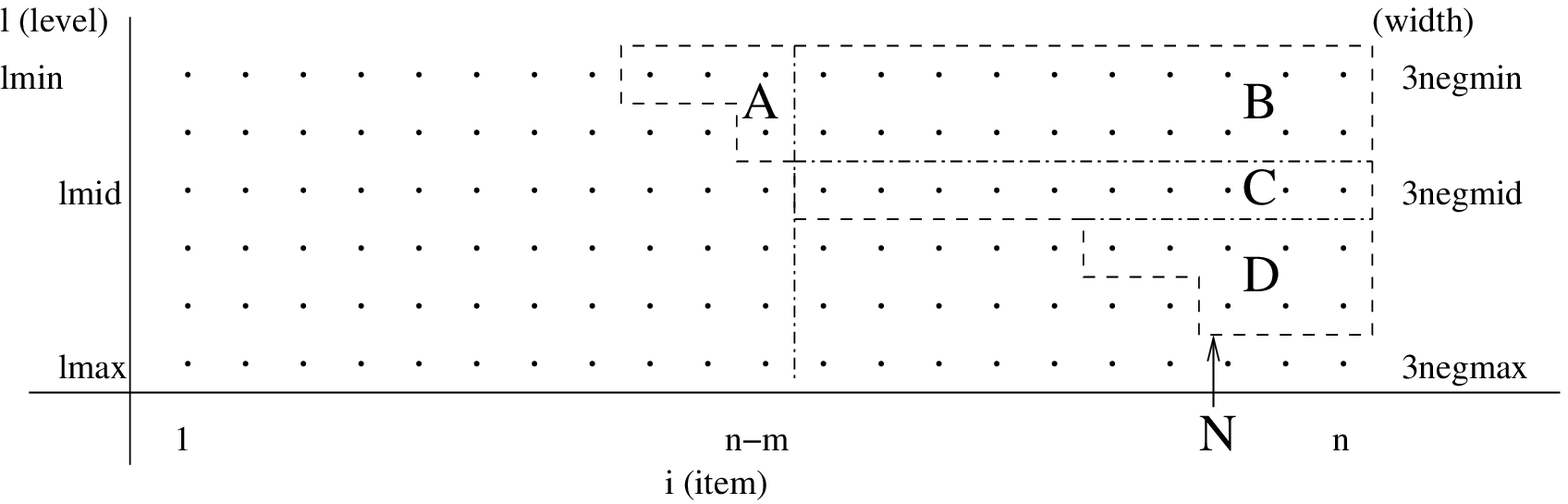}}
\caption{The set of nodes $I$, an optimal nodeset $N$, and disjoint subsets $\AlphaSet$, $\BetaSet$, $\GammaSet$, $\DeltaSet$}
\label{ABCD}
\end{figure*}

While not all optimal codes are monotonic, using the aforementioned
tie-breaking techniques, the algorithm always results in a monotonic
code, one that has minimum maximum length among all monotonic optimal
codes.  Examples of monotonic nodesets include the sets of nodes
enclosed by dashed lines in Fig.~\ref{nodesetnum} and Fig.~\ref{ABCD}.
In the latter case, $n = 21$, $\algd = 3$, $l_{\min} = 2$, and
$l_{\max} = 8$, so $\rhot = 2/3$.

In \cite{LaHi}, monotonicity allows trading off a constant factor of
time for drastically reduced space complexity for length-limited
binary codes.  We extend this to the bounded-length problem.
Note that the total width of items that are each less than or equal to
width $\rho$ is less than~$2n\rho$.  Thus, when we are processing
items and packages of width $\rho$, fewer than $2n$ packages are
kept in memory.  The key idea in reducing space complexity is to keep
only four attributes of each package in memory instead of the full
contents.  In this manner, we use $\order(n)$ space while retaining enough
information to reconstruct the optimal nodeset in algorithmic
postprocessing.

Package attributes allow us to divide the problem into two subproblems
with total complexity that is at most half that of the original
problem.  Define $$l_{\mid} \definedas \left\lfloor \frac{1}{2}
(l_{\max}+l_{\min}+1) \right\rfloor .$$ For each package $S$, we
retain only the following attributes:
\begin{enumerate}
\item $\mu(S) \das \sum_{(i,l) \in S} \mu(i,l)$
\item $\rho(S) \das \sum_{(i,l) \in S} \rho(i,l)$
\item $\nu(S) \das |S \cap I_{\mid}|$
\item $\psi(S) \das \sum_{(i,l) \in S \cap I_{\hit}} \rho(i,l)$
\end{enumerate}
where $I_{\hi} \definedas \{ (i,l)~|~l>l_{\mid} \}$ and $I_{\mid}
\das \{ (i,l)~|~l=l_{\mid} \}$.  We also define $I_\lo \das
\{(i,l)~|~l<l_{\mid} \}$.

With only these parameters, the ``first run'' of the algorithm takes
$\order(n)$ space.  The output of this run is the package attributes
of the optimal nodeset $N$.  Thus, at the end of this first run, we
know the value for $n_\nu \das \nu(N)$, and we can consider $N$ as the
disjoint union of four sets, shown in Fig.~\ref{ABCD}:
\begin{enumerate}
\item $\AlphaSet$ = nodes in $N \cap I_\lo$ with indexes in $[1,n-n_\nu]$,
\item $\BetaSet$ = nodes in $N \cap I_\lo$ with indexes in $[n-n_\nu+1,n]$,
\item $\GammaSet$ = nodes in $N \cap I_{\mid}$,
\item $\DeltaSet$ = nodes in $N \cap I_\hi$.
\end{enumerate}
Due to the monotonicity of $N$, it is clear that $\BetaSet =
[n-n_\nu+1,n] \times [l_{\min}+1, l_\mid-1]$ and $\GammaSet =
[n-n_\nu+1, n] \times \{l_\mid\}$.  Note then that $\rho(\BetaSet) =
(n_\nu) (\algd^{-l_{\min}}-\algd^{1-l_{\mid}})/(\algd-1)$ and
$\rho(\GammaSet) = n_\nu \algd^{-l_\midt}$.  Thus we need merely to
recompute which nodes are in $\AlphaSet$ and in~$\DeltaSet$.

Because $\DeltaSet$ is a subset of $I_{\hi}$, $\rho(\DeltaSet) =
\psi(N)$ and $\rho(\AlphaSet) = \rho(N) - \rho(\BetaSet) -
\rho(\GammaSet) - \rho(\DeltaSet)$.  Given their respective widths,
$\AlphaSet$ is a minimal weight subset of $[1,n-n_\nu] \times
[l_{\min}+1,l_{\mid}-1]$ and $\DeltaSet$ is a minimal weight subset of
$[n-n_\nu+1, n] \times [l_{\mid}+1,l_{\max}]$.  These will be
monotonic if the overall nodeset is monotonic.  The nodes at each
level of $\AlphaSet$ and $\DeltaSet$ can thus be found by recursive
calls to the algorithm.  This approach uses only $\order(n)$ space
while preserving time complexity as in \cite{LaHi}.  

There are changes we can make to the algorithm that, for certain
inputs, will result in even better performance.  For example, if
$l_{\max} \approx \log_\algd n$, then, rather than minimizing the
weight of nodes of a certain total width, it is easier to maximize
weight and find the complementary set of nodes.  Similarly, if most
input symbols have one of a handful of probability values, one can
consider this and simplify calculations.  These and other similar
optimizations have been done in the past for the special case
$\varphi(\delta)=\delta$, $l_{\min}=0$,
$\algd=2$\cite{KMT,LiMo,MTK,TuMo,TuMo2}, though we will not address or
extend such improvements here.

Note also that there are cases in which we can find a better upper
bound for codeword length than $l_{\max}$ or a better lower bound than
$l_{\min}$.  In such cases, complexity is accordingly reduced, and,
when $l_{\max}$ is effectively trivial (e.g., $l_{\max}=n-1$), and the
Package-Merge approach can be replaced by conventional (linear-time)
Huffman coding approaches.  Likewise, when $\varphi(\delta)=\delta$
and $l_{\max}-l_{\min}$ is not $\order(\log n)$, an approach similar
to that of \cite{LaPr1} as applied in \cite{Schi} has better
asymptotic performance.  These alternative approaches are omitted due
to space and can be found at \cite{Baer20}.

\section{Fringe-limited Prefix Coding}
\label{conclusion}

An important problem that can be solved with the techniques in this
paper is that of finding an optimal code given an upper bound on
fringe, the difference between minimum and maximum codeword length;
such a problem is proposed in \cite[p.~121]{Abr01}, where it is
suggested that if there are $b-1$ codes better than the best code with
fringe at most $d$, one can find this $b$-best code with the $\order(b
n^3)$-time algorithm in \cite[pp.~890--891]{AnHa}, thus solving the
fringe-limited problem.  However, this presumes we know an upper bound
for $b$ before running this algorithm.  More importantly, if a
probability vector is far from uniform, $b$ can be very large, since
the number of viable code trees is $\Theta(1.794\ldots^n)$\cite{KMN,
FlPr}.  Thus this is a poor approach in general.  Instead, we can use
the aforementioned algorithms for finding the optimal bounded-length
code with codeword lengths restricted to $[l'-d,l']$ for each $l' \in
\{\lceil \log_\algd n \rceil, \lceil \log_\algd n \rceil + 1, \ldots,
\lfloor \log_\algd n \rfloor + d\}$, keeping the best of these codes;
this covers all feasible cases of fringe upper bounded by~$d$.  (Here
we again assume, without loss of generality, that $n \bmod{(\algd-1)}
\equiv 1$.)  The overall procedure thus has time complexity
$\order(nd^2)$ and $\order(n)$ space complexity.

\section*{Acknowledgments}

The author wishes to thank Zhen Zhang for first bringing a related
problem to his attention and David Morgenthaler for constructive
discussions on this topic.

\ifx \cyr \undefined \let \cyr = \relax \fi

\end{document}